\documentclass[aps,prl,reprint,groupedaddress,showpacs]{revtex4-1}

\usepackage{amsmath}

\usepackage{amssymb}
\usepackage{amsthm}

\usepackage{braket}

\usepackage{graphicx}

\usepackage{color}

\begin{document}


\title{A general framework for microscopically reversible processes with memory}


\author{J. Ricardo Arias-Gonzalez}
\email[Corresponding author: ricardo.arias@imdea.org]{}
\affiliation{Instituto Madrile\~{n}o de Estudios Avanzados en Nanociencia,
C/Faraday 9, Cantoblanco, 28049 Madrid, Spain}
\affiliation{CNB-CSIC-IMDEA Nanociencia Associated Unit
``Unidad de Nanobiotecnolog\'{i}a"}


\date{\today}

\begin{abstract}
Statistical Mechanics deals with ensembles of microstates that are compatible
with fixed constraints and that on average define a thermodynamic macrostate.
The evolution of a small system is normally subjected to changing constraints
and involve a stochastic dependence on previous events.
Here, we develop a theory for reversible processes with memory that comprises
equilibrium statistics and that converges to the same physics in the limit
of independent events.
This framework is based on the characterization of single phase-space
pathways and is used to derive ensemble-average dynamics in stochastic systems
driven by a protocol in the limit of no friction.
We show that the state of a system depends on its history to the extent of
attaining a one-to-one correspondence between states and pathways when
memory covers all the previous events.
Equilibrium appears as the consequence of exploring all pathways that connect
two states by all procedures.
This theory is useful to interpret single-molecule experiments in Biophysics
and other fields in Nanoscience and an adequate platform for a general theory 
of irreversible processes.
\end{abstract}


\maketitle


Reversibility refers to quasistatic processes that invert isentropically.
Such processes involve a sufficiently slow dynamics to prevent heat flows,
more in depth, they take place through a large enough succession of states
along which there is not energy dissipation.
Reversible processes are normally analyzed by equilibrium statistics:
a so-called partition function describes the thermodynamic
properties of the system throughout all possible sequences of events
along infinite time intervals.
For systems whose fate is not dependent on either the past or the present,
like many macroscopic systems, both equilibrium and frictionless
quasistatic processes can be examined through the same mathematical framework
because the system is able to explore all the possible configurations of
states in a sufficiently long time.

For small systems~\cite{Bustamante2005}, namely, those for which the energy
exchanges are smaller or similar to the thermal level,
the system may only evolve
along one of the possible trajectories by a certain protocol.
In these cases, the balance of energy
in terms of work and heat as a function of the temperature
has to be evaluated at the single phase-space pathway level.
In addition, the protocol by which the system evolves has to be
considered when ensemble-average thermodynamics are addressed,
especially when memory effects are present~\cite{Arias-Gonzalez2014a}.
Such cases are common in the nanoscale. For many biophysical processes that
take place in the cell, the study of each molecular trajectory
individually becomes crucial for a complete comprehension of the role of
fluctuations~\cite{Ritort2008}.
Biophysical processes have traditionally
been analyzed by bulk (ensemble-average) strategies but the importance
of tackling them at the single-molecule level and at the single-chemical
reaction level has raised much both scientific and technological interest in
the last twenty years~\cite{Bustamante2008,Arias-Gonzalez2014}.
Replication, transcription and translation in Molecular Biology, just to name
a few, are processes whose thorough investigation requires single-molecule
approaches~\cite{Bustamante2011}:
nucleotides or aminoacids are incorporated
sequentially by a protein whose operation determines a
certain copying direction and a mechanism,
both of them responsible for chain stability and information fidelity.

It is therefore important to understand thermodynamic processes from a unified
point of view.
Here, we study reversible processes with memory and develop an general
framework for their analysis that comprises equilibrium and frictionless
quasistatic processes,
both of which considered from the single-pathway to the ensemble-average level.

The mechanism and external conditions that determines the protocol by which a
system evolves is
characterized by the so-called control parameter, $\lambda$,
which may actually be a set of parameters that describe
the state of the thermal bath and the constraints over the system.
For instance, in DNA replication, which to fix ideas is the example we will
refer to in this paper, the protocol is dictated, apart from other
environmental and chemical conditions,
by the DNA polymerase mechanism, which is stepwise from the 3' to the 5'
template end~\cite{Arias-Gonzalez2012}.
It is clear from this example that a protocol does not necessarily comprise
macroscopic variables that do not fluctuate, as has been normally
described~\cite{Bustamante2005,Ritort2008}.
In fact, microscopic processes in the cell (e.g., during metabolism)
are non-equilibrium without the need of invoking external,
non-fluctuating macroscopic variables that drive the system between
two states.

For the sake of clarity, from now on we will reserve the term
{\it protocol} for constraints that change with time and will denominate
{\it mechanism} (or {\it procedure}) the methodology (or operating mode)
that the system is allowed to employ (i.e., complying with the constraints)
to evolve.
In short, a protocol constrains the mechanisms by which a system can evolve;
hence, several mechanisms are compatible with the same protocol.

Given an initial and a final state of the system
at time instants $t=1$ and $t=n$, respectively,
every pathway that connects them will be specified by a temporal
sequence, $\nu=\{x_1,\ldots,x_t,\ldots,x_n\}$,
of stochastic states $X_t=x_t$ ($x_t \in {\cal X}$, being ${\cal X}$ the
alphabet of the random symbols or domain of the random variables, and
$t=1,\ldots,n$) under protocol $\lambda$.
The probability of a pathway is a function of its energy
$E_{\nu} = \sum_{t=1}^n E_t$,
which is a sum over the energies of the states that the system has passed
through in its evolution between states $x_1$ and $x_n$.
The energy of each state is in turn a function of the previous states, namely,
$E_t = E(x_t;x_{t-1},\ldots,x_1)$, which account for the memory,
i.e. the relative interactions of every present state, $x_t$, with its
previous ones, $\{x_{t-1},\ldots,x_1\}$.
The protocol determines not only the probability of the states but also the
number (or density for continuous random variables) 
of states that the system goes through,
i.e. protocols that connect $x_1$ and $x_n$ faster
involve a lower number (density) of intermediate states.

Within this scheme, the state of the system at time $t$ is not only
determined by the value of $X_t$ but also by how $x_t$ has been reached
because its energy depends on the sequence of previous events as
$E(x_t;x_{t-1},\ldots,x_1)$. Therefore, when there is confusion,
we will use the term {\it event} or {\it substate} for $x_t$ and reserve the
term state for the state itself plus its history.
We will further distinguish between {\it quasistate} and {\it state},
namely, a quasistate will be the ordered sequence of events until time $t$,
$\nu_t = \{x_1,\ldots,x_t\}$ ($\ket{\nu_t}$ in Quantum
Mechanics~\cite{Arias-Gonzalez2014a}), i.e. the substate at time $t$
($x_t$)
plus the ordered sequence of previous events $\{x_1,\ldots,x_{t-1}\}$,
and a state will be the ensemble-average over the pathways that the system can
follow until time $t$ driven by protocol $\lambda$.
A quasistate of a system is thus ultimately determined by its pathway and
viceversa when memory extends to the complete history of the system at
every time step.
A state in turn is soley determined by the protocol by which events have been
driven. In the limit of independent events, substates and quasistates are
equivalent and the term pathway is not necessary since quasistates do not
depend on how they have been reached by the
system (see the Independence Limit theorem~\cite{Arias-Gonzalez2014a}).
We will use the term {\it equilibrium state} to further refer to states
that have been reached by exploring all protocols and pathways,
i.e. to protocol- and pathway-independent states,
the rationale of which will become clearer later in this paper.
We will also show that in the limit of independent events,
states and equilibrium states refer to the same concept.

We will treat sequences $\nu$ as directional, stochastic chains with
memory~\cite{Arias-Gonzalez2014a} in the time domain.
It is important to note that the stochastic state of the system at each time
instant, $x_t$, may be well involve a set of stochastic variables according to
the degrees of freedom of the system. For example, in mechanical systems,
$x_t = \left ( q_h (t), p_h (t) \right )$, $h=1,\ldots,D$, where $D$ is the
number of degrees of
freedom and $q_h$ and $p_h$ are generalized space and momentum coordinates,
respectively. In information systems, both artificial and natural like
DNA replication, transcription or translation, $x_t$ is a set of symbolic
random variables addressing bits.

The DNA replication process actually comprises both a space- and time-dependent
directional, stochastic chain with memory~\cite{Arias-Gonzalez2012}.
More in depth, a DNA polymer, which is a material chain made up of
deoxyribonucleotide monophosphates (dNMP), grows directionally
as a function of the time. As explained in~\cite{Arias-Gonzalez2014a},
each incorporated dNMP at time $t$, $\tilde{x}_t$, can be conceived as the
outcome of a random variable, $\tilde{X}_t$, which probability is conditioned
by the previously incorporated dNMPs,
$\text{Pr} \{ \tilde{X}_t = \tilde{x}_t |
     \tilde{X}_{t-1} = \tilde{x}_{t-1},\ldots,\tilde{X}_1=\tilde{x}_1 \}$,
due to physical interactions.
A substate of the system, the growing DNA strand, can be described at
instant $t$ by the directionally ordered sequence of dNMPs,
$\tilde{\nu}_t=\{\tilde{x}_1,\ldots,\tilde{x}_{m(t)}\}$, where $m(t)$ is the
total number of dNMPs in the DNA chain at $t$.
Therefore, a pathway, $\nu$, above defined as a sequence of events, is for
this example a temporal succession of spatial sequences,
$\nu =\{\tilde{\nu}_1,\ldots,\tilde{\nu}_t,\ldots,\tilde{\nu}_n\}$.
Note that, while the cardinality of $\tilde{\nu}_t$ can be higher, equal or
lower than that of $\tilde{\nu}_{t+1}$,
i.e. $|\tilde{\nu}_t| \lessgtr |\tilde{\nu}_{t+1}|$,
because objects can be incorporated, removed or replaced as time progresses,
the cardinality of the temporal chain always grows,
$|\nu_t| = |\nu_{t+1}| - 1$, because time always increases
adding events to $\nu$.

The DNA replication example is of model significance because it shows how
physical interactions influence both the spatial arrangements and the
stochastic dependence of previous events in the history of the system. 
When the system under study is spatially more involved than a linear
arrangement of objects,
the material interactions may be more complex implying three-dimensional
interactions, likely over all spatial directions and neighbours,
but the time-dependence is always a directional, stochastic chain of events
because this coordinate advances only in one direction.

We extend next the concept of thermodynamic function to individual chains
that have been constructed by a reversible process.
We will assign to each single-chain thermodynamic function, ``$A$", a
pathway, $\nu$, and a protocol, $\lambda$,
by which that sequence has been assembled.
We will use the following notation: $A_{\nu}^{(\lambda)}$.
For the particular cases of stepwise construction from left to right and right
to left in material chains like DNA,
we may use $\lambda =D_+$ and $\lambda=D_-$,
respectively~\cite{Arias-Gonzalez2014a}.

We will constrain our analysis to the canonical and microcanonical ensembles
although the formalism can be easily extended to other statistical
ensembles~\cite{Chandler1987}.
The chains constructed under the same protocol can be treated by using
expected values:
\begin{equation}\label{eq:lpotential}
A^{(\lambda)} \equiv \left \langle A^{(\lambda)}_{\nu} \right \rangle_{\lambda}
= \sum_{\nu=1}^N p_{\nu}^{(\lambda)} A_{\nu}^{(\lambda)},
\end{equation}
\noindent
where $p_{\nu}^{(\lambda)}$ is defined according to protocol $\lambda$ by
using the corresponding sequence-dependent partition
function~\cite{Arias-Gonzalez2014a}:
\begin{equation}\label{eq:lprob}
p_{\nu}^{(\lambda)} = \frac{e^{-\beta E_{\nu}}}{Z_{\nu}^{(\lambda)}},
\end{equation}
\noindent
such that $\sum_{\nu=1}^N p_{\nu}^{(\lambda)} =1$, where $\beta = 1/ k T$
being $T$ the temperature and $k$ the Boltzmann constant.
As a directional chain~\cite{Arias-Gonzalez2014a},
the partition function $Z_{\nu}^{(\lambda)}$ has to be
evaluated according to each protocol $\lambda$ and with respect to a certain
pathway $\nu$, i.e. calculating individual
probabilities according to the available configurations at each time step:
\begin{eqnarray}\label{eq:Znu2}
Z_{\nu}^{(\lambda)} & \equiv &
\sum_{\nu'(\lambda)=1}^{N} \exp{\left( -\beta  E_{\nu' \nu} \right)},
\end{eqnarray}
\noindent
where $E_{\nu' \nu}$ is the two-sequence energy~\cite{Arias-Gonzalez2014a}:
\begin{equation}\label{eq:2seqEnergy}
E_{\nu' \nu} \equiv \sum_{t=1}^{n} E\left( x'_t; x_{t-1}, \ldots, x_1 \right),
\end{equation}
\noindent
$N= |{\cal X}|^n$ is the number of configurations, which is the result of
combining $n$ events and $|{\cal X}|$ possibilities for each event, and
subindex $\nu'(\lambda)$ in the sigma symbol reminds that the sum
over the multiple $x'_t$ variables, which are correlated due to memory effects,
has to be evaluated according to the constraints imposed by the protocol.
For example, a material one-dimensional chain may be constructed by
incorporating objects on a one-by-one basis and directionally,
either $\lambda=D_+$ or $\lambda=D_-$;
it can also be constructed, e.g. by incorporating more than one object
at a time and/or by alternating senses at each step, even by including
editions, which implies removing objects.
Typical copying systems, either natural like DNA replication or artificial like
tape-based technologies, generate copies stepwisely in one sense (say $D_+$)
and corrections by removing symbols in the opposite direction ($D_-$).

The equilibrium Statistical Physics is formulated by using the standard
partition function, namely
\begin{equation}\label{eq:ProbIsing}
p_{\nu} = \frac{\exp{\left( -\beta E_{\nu} \right)}}{Z}, \,\,\,\,\,\,\,
Z= \sum_{\nu=1}^{N} \exp{\left( -\beta E_{\nu} \right)}.
\end{equation}
\noindent
Partition function $Z$ does not
make any assumptions on a particular protocol and therefore it comprises all
the possibilities for all the protocols~\cite{Arias-Gonzalez2014a}.
In fact, sequence-dependent and equilibrium partition functions fulfill next
relations:
\begin{eqnarray}\label{eq:meanl1Z}
      \left\langle \frac{1}{Z_{\nu}^{(\lambda)}} \right\rangle = \frac{1}{Z},
\\ [+1mm]
\label{eq:meanlZ}
      \left\langle Z_{\nu}^{(\lambda)} \right\rangle_{\lambda} = Z,
\end{eqnarray}
\noindent
which are trivially demonstrated from the definition of
$p_{\nu}$ and $p_{\nu}^{(\lambda)}$~\cite{Arias-Gonzalez2014a}.
These equations are valid for every protocol $\lambda$, thus indicating that
the equilibrium partition function is an average over all possible sequences
for a fixed protocol, independently on which the protocol is. In other
words, $\left\langle p_{\nu}^{(\lambda)}/p_{\nu} \right\rangle =1$ and
$\left\langle p_{\nu}/p_{\nu}^{(\lambda)} \right\rangle_{\lambda} =1$.

The equilibrium partition function $Z$ measures the
number of available pathways $\nu$ that connect the states at $t=1$ and $t=n$,
constructed as temporal sequences of
stochastic events $X_t$ and statistically
weighted by their energies $\sum_t E_t$, which account for the memory,
i.e. the relative interactions of every present substate, $x_t$, with its
previous ones, $\{x_1,\ldots,x_{t-1}\}$.
The sequence-dependent partition function, $Z_{\nu}^{(\lambda)}$, in contrast,
measures the energy-weighted number of pathways that connect these states
considering that at each step, $x_t$, the previous events,
$\{x_{t-1},\ldots,x_1\}$, are unchangeable and that the sequence of events is
stochastically determined by protocol $\lambda$.
In short, $Z_{\nu}^{(\lambda)}$ measures the energy-weighted number
of available pathways that connect two states
when the history and the driving protocol are fixed.
In fact, the probability for sequence-dependent statistics fulfills
\begin{equation}\label{eq:PseqPartit}
p_{\nu_{t+1}}^{(\lambda)} =
 p_{\nu_t}^{(\lambda)} \times
p^{(\lambda)} \left( x_{t+1} | x_t,\ldots,x_1 \right),
\end{equation}
\noindent
where
$p^{(\lambda)} \left( x_{t+1} | x_t,\ldots,x_1 \right) =
\exp \left( -\beta E_{t+1} \right) / Z_{t+1}^{(\lambda)}$
and
\begin{equation}\label{eq:ZseqPartit}
Z_{t+1}^{(\lambda)} = \sum_{x'_{t+1} \in {\cal X}}
\exp \left( -\beta E \left (x'_{t+1};x_t,\ldots,x_1 \right) \right),
\end{equation}
being
$Z_{\nu}^{(\lambda)} = \prod_{t=1}^n Z_t^{(\lambda)}$,
which is a consequence of the fact that probabilities are stepwisely
constructed with fixed history as time progresses.
While equilibrium probabilities can be formally expressed like 
in Eq.~(\ref{eq:PseqPartit}), the sums in the partition function
$Z = \prod_{t=1}^n Z_t$ are
nested thus not rendering a product of independent factors
$Z_t=\sum_{x_t \in {\cal X}}
\exp \left( -\beta E \left (x_t;x_{t-1},\ldots,x_1 \right) \right)$
like in Eq.~(\ref{eq:ZseqPartit})~\cite{Arias-Gonzalez2012,Arias-Gonzalez2014a}.

We can now define the thermodynamic potentials, ``$U$", ``$F$" and ``$S$", for
single trajectories, namely, the pathway- and protocol-dependent
Internal Energy, Helmholtz Free Energy and Entropy:
\begin{eqnarray}\label{eq:GenTherPot1}
U_{\nu}^{(\lambda)} \equiv E_{\nu},
\\ [+1mm] \label{eq:GenTherPot2}
F_{\nu}^{(\lambda)} \equiv -kT\ln Z_{\nu}^{(\lambda)},
\\ [+1mm] \label{eq:GenTherPot3}
S_{\nu}^{(\lambda)} \equiv -k \ln p_{\nu}^{(\lambda)},
\end{eqnarray}
\noindent
fulfilling
\begin{equation}\label{eq:Econserv}
F_{\nu}^{(\lambda)} = U_{\nu}^{(\lambda)} - TS_{\nu}^{(\lambda)},
\end{equation}
which is the energy conservation. These potentials characterize the quasistates
of the system at time $t=n$.

Note that $E_{\nu}$ is independent of the protocol but not
$F_{\nu}^{(\lambda)}$ or $S_{\nu}^{(\lambda)}$.
These functions can be understood from a microcanonical
point of view as the thermodynamic potentials for fixed energy
$E_{\nu}$. When memory is extended to the complete history of the system,
there is in general a one-to-one relationship between pathways and
energies but when memory is limited to a finite number of previous events,
a degeneration of pathways with the same energy $E_{\nu}$ arises,
what renders statistical meaning to the entropy within the framework of the
microcanonical ensemble. Namely,
the entropy can be expressed in the form of Boltzmann formula,
``$S=k\ln \omega$",
by identifying $\omega_{\nu}^{(\lambda)}=1/p_{\nu}^{(\lambda)}$.
When memory is finite, $S$ is the entropy of the system at
time $t=n$ considering that it has evolved to this state by pathways that
result from an ensemble of configurations of $n$ stochastic substates with
equal energy $E_{\nu}$.
In constrat, $S_{\nu}^{(\lambda)}$ is the entropy of the system at
time $t=n$ considering that the available pathways with equal energy have fixed
history at each step and are traversed under a defined protocol ($\lambda$).

The ensemble-average thermodynamic potentials can be constructed
by taking expected values (see Eq.~(\ref{eq:lpotential}))
on Eqs.~(\ref{eq:GenTherPot1})-(\ref{eq:GenTherPot3}):
\begin{eqnarray}\label{eq:SeqDepTherPot1}
U^{(\lambda)} \equiv \left \langle E_{\nu} \right \rangle_{\lambda},
\\ [+1mm] \label{eq:SeqDepTherPot2}
F^{(\lambda)} \equiv -kT \left \langle \ln Z_{\nu}^{(\lambda)}
\right \rangle_{\lambda},
\\ [+1mm] \label{eq:SeqDepTherPot3}
S^{(\lambda)} \equiv -k \left \langle \ln p_{\nu}^{(\lambda)}
\right \rangle_{\lambda}.
\end{eqnarray}
\noindent
$U^{(\lambda)}$, $F^{(\lambda)}$ and $S^{(\lambda)}$ do not depend on the
pathway but they do on the protocol. Therefore, they will appear under the
name of protocol-dependent Internal Energy, Helmholtz Free Energy and Entropy,
respectively. These potentials characterize the states of the system
at time $t=n$.

The energy conservation for ensemble-average phenomena can also be expressed
in terms of protocol-dependent potentials as
\begin{equation}\label{eq:dSUF}
F^{(\lambda)} = U^{(\lambda)} - TS^{(\lambda)},
\end{equation}
\noindent
which arise by formally taking expected values on
Eq.~(\ref{eq:Econserv}).

\begin{proof}
From Eqs.~(\ref{eq:ProbIsing}), it follows that
\begin{eqnarray}\label{eq:demodSUF}
 S^{(\lambda)} & = &
-k \sum_{\nu=1}^N p^{(\lambda)}_{\nu} \ln p^{(\lambda)}_{\nu} =
-k \sum_{\nu=1}^N p^{(\lambda)}_{\nu} \left( -\beta E_{\nu} -
\ln Z_{\nu}^{(\lambda)}\right)
\nonumber \\ [+1mm]
& = &
k \left(
\beta \left\langle E_{\nu} \right\rangle_{\lambda} +
\left\langle \ln Z_{\nu} \right\rangle_{\lambda} \right)
= k \beta \left( U^{(\lambda)} - F^{(\lambda)} \right),
\nonumber
\end{eqnarray}
\noindent
which proves Eq.~(\ref{eq:dSUF}).
\end{proof}

It is easy to see that equilibrium thermodynamics is a particular
case of the above formalism. Namely, using
Eqs.~(\ref{eq:ProbIsing}) for the partition function and the probabilites,
the energy conservation, $F=U-TS$, appears naturally by dropping the
protocol superindex $\lambda$ on
Eqs.~(\ref{eq:SeqDepTherPot1})-(\ref{eq:SeqDepTherPot3}).
Potentials $U$, $F$ and $S$ characterize the equilibrium states of the system.

Noteworthy, thermodynamic functions $A_{\nu}^{(\lambda)}$,
which characterize the quasistates of a system, and $A^{(\lambda)}$, which
characterize the states of the system,
are different from the equilibrium state functions, $A$, which
are independent of time and of both protocol and pathway.

The internal energy for directional, stochastic chains fulfills:
\begin{equation}\label{eq:UD}
U^{(\lambda)} = - \left \langle
\frac{\partial}{\partial \beta} \ln Z_{\nu}^{(\lambda)}
\right \rangle_{\lambda}.
\end{equation}

\begin{proof}
From Eqs.~(\ref{eq:SeqDepTherPot1}) and~(\ref{eq:lpotential}),
\begin{eqnarray}\label{eq:demoUD}
\langle E_{\nu} \rangle_{\lambda} & = & \sum_{\nu=1}^N p^{(\lambda)}_{\nu}
E_{\nu} =
\sum_{\nu=1}^N \frac{e^{-\beta E_{\nu}}}{Z_{\nu}^{(\lambda)}} E_{\nu}
\nonumber \\ [+1mm]
& = &
-\sum_{\nu=1}^N \frac{1}{Z_{\nu}^{(\lambda)}}
\frac{\partial}{\partial\beta} e^{-\beta E_{\nu}} =
-\sum_{\nu=1}^N \frac{\partial}{\partial\beta} p^{(\lambda)}_{\nu}
\nonumber \\ [+1mm]
& - &
\sum_{\nu=1}^N p^{(\lambda)}_{\nu} \frac{\partial}{\partial\beta}
\ln Z_{\nu}^{(\lambda)}
= - \left \langle
\frac{\partial}{\partial \beta} \ln Z_{\nu}^{(\lambda)}
\right \rangle_{\lambda},
\nonumber
\end{eqnarray}
\noindent
which proves Eq.~(\ref{eq:UD}).
\end{proof}

Likewise, the entropy for directional, stochastic chains follows the next law:
\begin{equation}\label{eq:SD}
S^{(\lambda)} = - \left \langle
\frac{\partial}{\partial T} F_{\nu}^{(\lambda)}
\right \rangle_{\lambda}.
\end{equation}

\begin{proof}
From Eqs.~(\ref{eq:dSUF}),~(\ref{eq:SeqDepTherPot2}),~(\ref{eq:UD})
and~(\ref{eq:GenTherPot2}), in this order, it follows that
\begin{eqnarray}\label{eq:demoSD}
S^{(\lambda)} & = & -\frac{1}{T} \left ( F^{(\lambda)} - U^{(\lambda)} \right)
\nonumber \\ [+1mm]
& = &
-\frac{1}{T} \left \langle
-kT \ln Z_{\nu}^{(\lambda)}
+ \frac{\partial}{\partial \beta} \ln Z_{\nu}^{(\lambda)}
\right \rangle_{\lambda}
\nonumber \\ [+1mm]
& = &
- \left \langle
\frac{\partial}{\partial T} F_{\nu}^{(\lambda)} \right \rangle_{\lambda},
\nonumber
\end{eqnarray}
\noindent
which proves Eq.~(\ref{eq:SD}).
\end{proof}

Equations~(\ref{eq:UD}) and~(\ref{eq:SD}), while inmmediate in their
demonstrations, are not obvious. Their equilibrium analogues appear,
respectively, as particular cases within this formalism:
\begin{eqnarray}\label{eq:U}
U & \equiv & \langle E \rangle = - \left \langle
\frac{\partial}{\partial \beta} \ln Z \right \rangle =
-\frac{\partial}{\partial \beta} \ln Z,
\\ [+1mm] \label{eq:S}
S & \equiv & -k \langle \ln p_{\nu} \rangle =
- \left \langle \frac{\partial}{\partial T} F \right \rangle =
-\frac{\partial}{\partial T} F.
\end{eqnarray}

We next consider inequalities between sequence-dependent and equilibrium
Thermodynamics.
The Kullback-Leibler distances are positive, as expected:
\begin{eqnarray}\label{eq:KL1}
D \left( p_{\nu} || p_{\nu}^{(\lambda)} \right) = 
\left \langle \ln \frac{Z_{\nu}^{(\lambda)}}{Z} \right \rangle \geq 0,
\\ [+1mm]
\label{eq:KL2}
D \left( p_{\nu}^{(\lambda)} || p_{\nu} \right) = 
\left \langle \ln \frac{Z}{Z_{\nu}^{(\lambda)}} \right \rangle_{\lambda} \geq 0,
\end{eqnarray}
\noindent
where we have applied Jensen's inequality~\cite{Cover1991}
and Eqs.~(\ref{eq:meanl1Z}) and~(\ref{eq:meanlZ}).
Concerning the ensemble-average thermodynamic
functions, the following inequalities hold:
\begin{eqnarray}\label{eq:Fineq}
F^{(\lambda)} \geq F \equiv -\frac{1}{\beta} \ln Z,
\\ [+1mm]
\label{eq:Sineq}
S - S^{(\lambda)} \geq \frac{1}{T} \left( U - U^{(\lambda)} \right).
\end{eqnarray}
\noindent
Inequation~(\ref{eq:Fineq}) follows from the application of Jensen's
inequality~\cite{Cover1991} to $-\ln x$, which is a convex function of $x$.
The fact that $F^{(\lambda)} - F \geq 0$ physically means that a final
substate $x_n$ achieved under statistical equilibrium (i.e. by going through
all pathways under all possible protocols) is always more stable than when
the same state has been achieved under all the possible pathways but 
defined by only one specific protocol $\lambda$.
In fact, this inequality indicates that the system at $x_n$ can still achieve
statistical equilibrium by completing all pathways by the rest
of protocols, hence visiting all the configurations surrounding $x_n$ after
waiting for a sufficiently long time.  

Inequation~(\ref{eq:Sineq}) derives both from Ineq.~(\ref{eq:Fineq})
and from Eq.~(\ref{eq:meanlZ}).
The entropy of a system at substate $x_n$ can increase or decrease with
respect to the statistical equilibrium entropy value
(i.e., $S - S^{(\lambda)} \lessgtr 0$),
it depends on the protocol $\lambda$.
In fact, certain protocols can decrease the entropy largely at the cost of
high dissipations under non-equilibrium conditions~\cite{Andrieux2008b}.
Inequation~(\ref{eq:Sineq}) guarantees that if the
ensemble-average internal energy of the final substate $x_n$ achieved under
protocol $\lambda$ is lower than that achieved under statistical equilibrium
(namely, if $U - U^{(\lambda)} \geq 0$),
the final entropy decreases below the statistical equilibrium value
through this protocol (namely, then $S - S^{(\lambda)} \geq 0$).

From the ensemble-average energy conservation, Eq.~(\ref{eq:dSUF}), it is
clear that $U \geq F$ and that $U^{(\lambda)} \geq F^{(\lambda)}$, which mean
that the useful energies, $F$ or $F^{(\lambda)}$, are always lower than the
total energies, $U$ or $U^{(\lambda)}$, respectively, due to the entropic
term.
From these inequalities it follows that
\begin{equation}\label{eq:IneqUlF}
U^{(\lambda)} \geq F,
\end{equation}
which sets a minimal boundary for the ensemble-average internal energy of the
system at final substate $x_t$.
Former results on DNA replication in the reversible
limit~\cite{Arias-Gonzalez2012} are compatible with
inequalities~(\ref{eq:Fineq})-(\ref{eq:IneqUlF}).
In particular, the statistical equilibrium entropy of the
stochastic chain was lower than the entropy achieved under a directional
construction protocol and the same trend was observed for the internal energy,
both of them to a strength compatible with Ineq.~(\ref{eq:Sineq}) at the
temperature of that study.

From now on, we define the Hamiltonian of the system as
$H(\nu,  \alpha) = H(x_1,\ldots,x_n;\alpha_1,\ldots,\alpha_k)$, where
$\alpha_i$ are external parameters. If one of these parameters
experiences a small change, the system on average will oppose to this change
with a conjugate force $\gamma_i$ according to:
\begin{equation}\label{eq:ConjugForce}
\gamma_i^{(\lambda)} \equiv - \left \langle
\frac{\partial H(\nu; \alpha)}{\partial \alpha_i} \right \rangle_{\lambda} =
\frac{1}{\beta} \left \langle
\frac{\partial}{\partial \alpha_i} \ln Z_{\nu}^{(\lambda)}(\alpha)
\right \rangle_{\lambda},
\end{equation}
\noindent
which reduces to the well-known expression
$\gamma_i = \frac{1}{\beta} \frac{\partial \ln Z(\alpha)}{\partial \alpha_i}$
in equilibrium. $\gamma_i^{(\lambda)}$ can be interpreted in turn as the
ensemble-average force,
$\gamma_i^{(\lambda)} = \left \langle \gamma_{i, \nu}^{(\lambda)}
\right \rangle_{\lambda}$,
over the pathway-dependent forces,
$\gamma_{i, \nu}^{(\lambda)} \equiv
\frac{1}{\beta} \frac{\partial} {\partial \alpha_i}
\ln Z_{\nu}^{(\lambda)} (\alpha)$,
which represents the conjugate force for a small change
in $\alpha_i$ when the system follows pathway $\nu$ under protocol $\lambda$.

\begin{proof}
The demonstration of Eq.~(\ref{eq:ConjugForce}) is similar to that of the
internal energy, Eq~(\ref{eq:UD}):
\begin{eqnarray}\label{eq:demoConjugForce}
\gamma_i^{(\lambda)} & = & - \sum_{\nu=1}^N p_{\nu}^{(\lambda)}
\frac{\partial H(\nu; \alpha_i)}{\partial \alpha_i} = \frac{1}{\beta}
\sum_{\nu=1}^N \frac{1}{Z_{\nu}^{(\lambda)}}
\frac{\partial}{\partial \alpha_i} e^{-\beta H}
\nonumber \\ [+1mm]
& = &
\frac{1}{\beta} \sum_{\nu=1}^N \frac{\partial}{\partial \alpha_i}
p_{\nu}^{(\lambda)}+
\frac{1}{\beta} \sum_{\nu=1}^N p_{\nu}^{(\lambda)}
\frac{\partial}{\partial \alpha_i} \ln Z_{\nu}^{(\lambda)}
\nonumber \\ [+1mm]
& = &
\frac{1}{\beta} \left \langle
\frac{\partial}{\partial \alpha_i} \ln Z_{\nu}^{(\lambda)}
\right \rangle_{\lambda},
\nonumber
\end{eqnarray}
\noindent
where we have used $\partial T/\partial \alpha_i = 0$ because $\alpha_i$ are
external parameters different from $T$.
\end{proof}

Next, we derive the expression of the {\it equipartition theorem}.
To do this, we need to suppose that $x_i$ are continuous random variables
because, otherwise, it is not possible to think about small changes (it does
not make sense discussing about small changes of symbols or discrete 
variables). 
Let $x_i$ be a generalized position, $q_i$, or momentum, $p_i$. Then, the
equipartition theorem is
\begin{equation}\label{eq:EquipTh}
\left \langle x_i \frac{\partial H(\nu)}{\partial x_j} \right \rangle_{\lambda}
= kT \delta_{ij} - kT \left \langle x_i \frac{\partial}{\partial x_j}
\ln Z_{\nu}^{(\lambda)} \right \rangle_{\lambda},
\end{equation}
\noindent
which differs from the equilibrium equipartition theorem in the second term,
non-zero even when $i \neq j$. This term manifests the effect
of the interactions when the sequence is constructed under a protocol.
Certainly, the protocol limits the range of a degree of freedom (note the
minus sign between the first (equilibrium) and second term) because the
system is stochastically forced to follow certain pathways.
The interaction term becomes zero in the equilibrium limit since on average
there is no net opposition for the system to visit all substates by all
protocols.

\begin{proof}
For the sake of simplicity, we switch to continuous variables hence replacing
sums by integrals, following the demonstration found in Ref.~\cite{Pathria2011}
for the equilibrium equipartition theorem.
\begin{eqnarray}\label{eq:demoEquipTh}
\left \langle x_i \frac{\partial H(\nu)}{\partial x_j} \right \rangle_{\lambda}
& = &
\int d\nu p_{\nu}^{(\lambda)} x_i \frac{\partial H(\nu)}{\partial x_j}
\nonumber \\ [+1mm]
& = &
- \frac{1}{\beta}\int d\nu \frac{x_i}{Z_{\nu}^{(\lambda)}}
\frac{\partial}{\partial x_j}
e^{-\beta H(\nu)}
\nonumber \\ [+1mm]
& = &
- \frac{1}{\beta} \int d\nu \frac{\partial}{\partial x_j}
\left( \frac{x_i}{Z_{\nu}^{(\lambda)}} e^{-\beta H(\nu)} \right)
\nonumber \\ [+1mm]
& + &
\frac{1}{\beta} \int d\nu e^{-\beta H(\nu)} \frac{\partial}{\partial x_j}
\frac{x_i}{Z_{\nu}^{(\lambda)}},
\end{eqnarray}
where we have integrated over $x_j$ by parts. The first integral becomes:
\begin{equation}\label{eq:demoEquipTh1}
\int d\nu(j) \left[ -\frac{1}{\beta} x_i p_{\nu}^{(\lambda)}
\right]^{(x_j)_2}_{(x_j)_1}=0,
\end{equation}
\noindent
where $d\nu(j)$ denotes $d\nu$ devoid of $dx_j$, that is,
$d\nu \equiv d\nu(j)\times dx_j$. This integral vanishes because
$(x_j)_2$ and $(x_j)_1$ are extreme values of $x_j$, at which the Hamiltonian
becomes infinite~\cite{Pathria2011}.
At this step, special care must be taken with $Z_{\nu}^{(\lambda)}$, which,
unlike $Z$, is not a constant.
However, $Z_{\nu}^{(\lambda)} > 0$ because, in general,
the directional
partition function involves a sum over other $\exp-\beta H$ terms
that are not evaluated at their extreme values~\cite{Arias-Gonzalez2014a}.
The second integral in Eq.~(\ref{eq:demoEquipTh}) becomes
\begin{eqnarray}\label{eq:demoEquipTh2}
& & \frac{1}{\beta} \int d\nu e^{-\beta H(\nu)}
\left( \frac{\delta_{ij}}{Z_{\nu}^{(\lambda)}}
-\frac{x_i}{\left ( Z_{\nu}^{(\lambda)} \right ) ^2}
\frac{\partial Z_{\nu}^{(\lambda)}}{\partial x_j}\right)
\nonumber \\ [+1mm]
& = &
kT \delta_{ij} - kT \int d\nu p_{\nu}^{(\lambda)}
x_i \frac{\partial}{\partial x_j}
\ln Z_{\nu}^{(\lambda)}
\end{eqnarray}
\noindent
which proves Eq.~(\ref{eq:EquipTh}).
\end{proof}

Finally, we will derive an expression that relates the sequence-dependent
partition function with the respective sequence-dependent energies,
which can be useful from an experimental point of view.
For each sequence, $\nu$, the sequence-dependent partition function can be
estimated if knowledge of the energies $E_{\mu \nu}$ exists through the next
relation:
\begin{equation}\label{eq:ZnuEstim}
\frac{Z_{\nu}^{(\lambda)}}{Z} =
\left\langle  e^{-\beta \left( E_{\mu \nu} - E_{\mu}\right)} \right\rangle,
\end{equation}
\noindent
where the expected value is taken over all sequences
$\mu \equiv \mu (\lambda) = 1,\ldots,N$
with the equilibrium probability distribution
$p_{\mu}$, Eq.~(\ref{eq:ProbIsing}).
It is in general difficult to apply Eq.~(\ref{eq:ZnuEstim}) when
interactions extend over many neighbours because energies $E_{\mu \nu}$ involve
many combinations for a defined protocol. When sufficient
knowledge on the system is gathered, like for example in DNA replication,
it is possible to restrict the elements of the
energy data set~\cite{Arias-Gonzalez2012,SantaLucia2004}.

\begin{proof}
From Eqs.~(\ref{eq:Znu2}) and~(\ref{eq:ProbIsing}), it is straightforward that
\begin{eqnarray}\label{eq:demoZnuEstim}
\frac{Z_{\nu}^{(\lambda)}}{Z} & = & \frac{Z_{\nu}^{(\lambda)}-Z}{Z} + 1 =
\frac{1}{Z} \sum_{\mu(\lambda)=1}^N \left(e^{-\beta E_{\mu \nu}} -
e^{-\beta E_{\mu}}\right) +1
\nonumber \\ [+1mm]
& = &
\frac{1}{Z} \sum_{\mu(\lambda)=1}^N e^{-\beta E_{\mu}}
\left( e^{-\beta \left(E_{\mu \nu}-E_{\mu}\right)}-1\right) + 1
\nonumber \\ [+1mm]
& = &
\sum_{\mu(\lambda)=1}^N p_{\mu}
\left( e^{-\beta \left(E_{\mu \nu}-E_{\mu}\right)}-1\right) +1 =
\left\langle  e^{-\beta \left( E_{\mu \nu} - E_{\mu}\right)} \right\rangle
\nonumber
\end{eqnarray}
\end{proof}

We have extended Thermodynamics to states attainable under specific sequences
of events of a system driven by changing constraints in the frictionless limit.
To do that, we have introduced pathway- and protocol-dependent functions,
including thermodynamic potentials, that
characterize the states of the system in the presence of memory.
The canonical ensemble
---with fixed temperature, $T$, and number (or density) of
events, $n$, along the pathways that connect two states under a protocol---
uses protocol-dependent potentials whereas the microcanonical
ensemble
---with fixed pathway energy, $E_{\nu}$, besides $T$ and $n$---
uses both pathway- and protocol-dependent potentials for the thermodynamic
characterization of a pathway described by the system under a protocol.

Although contemplated within the same mathematical framework,
our theory discriminates between reversible and equilibrium thermodynamics,
which converge to the same physics when memory effects become negligible.
We find that a system attains an equilibrium state when it transforms into it
after tackling all pathways and protocols. The thermodynamic functions become
so-called (equilibrium) state functions,
that is, pathway and protocol-independent, in these conditions.

We have defined substates (or events) as the stages that the system goes
through in its evolution along a pathway, and have considered quasistates as
substates with history.
We have reserved the term state to what is attained by a system after visiting
all the quasistates compatible with fixed constraints, and equilibrium state
when quasistates are visited without constraints.
When memory is present, the number
of quasistates reached by a system increases with the number of events that the
system recalls because each quasistate involves a configurational history of
substates.
In fact, when memory extends to all previous events at each time step
there is a bijection between quasistates and pathways;
the system actually restarts whenever it explores new pathways in these
conditions.
If memory effects can be cut off down to a finite number of previous events,
as for example in Markov (memoryless) dynamics or in the case of independent
events, quasistates can be recurrently visited within a particular pathway,
that is, without restarting.
The lower the number of nearest
temporal neighbours to be considered in the memory, the lower the revisiting
period. This revisiting period can be assumed as the so-called Poincar\'e
recurrence time, which increases with the number of past events
that stochastically influence the present.
In the limit in which the origin of the system is $t=-\infty$ and the memory
extends to all previous events, the Poincar\'e recurrence time tends
to infinity because the system cannot be restarted.

The evolution of a system is a consequence of the existence of a protocol,
which represents constraints that change with time.
If the protocol is sufficiently smooth in the time dependence, the system
evolves sufficiently slowly to visit many substates by virtually all the
fast enough mechanisms, thus approaching an
infinite succession of equilibrium states ($n \to \infty$).
On the contrary, if the protocol has a sharp time dependence, the constraints
are too strong for the system to visit a sufficient number of substates to 
reach equilibrium ($n < \infty$) and
hence its evolution is more dramatically marked by the protocol;
in these conditions, the system evolves away from equilibrium.
An equilibrium transformation is therefore an idealization, which is approached
by protocols that allow the system to visit a statistically representative
(sufficiently large) number of substates by many procedures
while it changes its ensemble-average state (or macrostate),
as defined by its characteristic protocol-dependent thermodynamic potentials.

The existence of a protocol thefore limits the mechanisms
that operate and the number of substates that are required to
drive the system between two quasistates.
Consequently, ensemble averages and time averages
are not interchangeable, which make evolutions no longer ergodic.
Since equilibrium and reversible Thermodynamics are essentially the same
when neither the present nor the past of the system influence the future,
the existence of a protocol is only significant in the presence of memory
(that is, for both so-called Markovian and non-Markovian
processes~\cite{Breuer2012,Rivas2014}).

Considering that Fluctuation Theorems assume both
microscopic reversibility and Markovianity~\cite{Bustamante2005,Ritort2008},
we anticipate that our theory can describe Non-equilibrium Thermodynamics
within a unified framework.

\noindent
\textbf{Acknowledgments}\\
Work supported by IMDEA Nanociencia.


\begin{thebibliography}{14}
\bibitem{Bustamante2005} Bustamante, C., Liphardt, J. \& Ritort, F. The nonequilibrium thermodynamics of small systems. \textit{Physics Today}, \textbf{58}, 43-48 (2005).

\bibitem{Arias-Gonzalez2014a} Arias-Gonzalez, J.~R. Statistical physics of directional, stochastic chains with memory. arXiv:1511.06139 [cond-mat.stat-mech]

\bibitem{Ritort2008} Ritort, F. Nonequilibrium fluctuations in small systems: From physics to biology. In: Rice, S.~A., editor. \textit{Adv. Chem. Phys.}, Wiley. Chapter 2, \textbf{137}, 31-123 (2008).

\bibitem{Bustamante2008} Bustamante, C. In singulo Biochemistry: When Less Is More. \textit{Annu. Rev. Biochem.}, \textbf{77}, 45-50 (2008).

\bibitem{Arias-Gonzalez2014} Arias-Gonzalez, J.~R. Single-molecule portrait of DNA and RNA double helices. \textit{Integr. Biol.} \textbf{6}, 904-925 (2014).

\bibitem{Bustamante2011} Bustamante, C., Cheng, C. \& Mejia, Y.~X. Revisiting the Central Dogma One Molecule at a Time. \textit{Cell}, \textbf{144}, 480-497 (2011).

\bibitem{Arias-Gonzalez2012} Arias-Gonzalez, J. R. Entropy involved in fidelity of DNA replication. \textit{PLoS ONE}, \textbf{7}, e42272 (2012).

\bibitem{Chandler1987} Chandler, D. {\it Introduction to Modern Statistical Mechanics} (Oxford University Press, 1987).

\bibitem{Cover1991} Cover, T.~M. \& Thomas, J.~A. {\it Elements of Information Theory} (John Wiley \& Sons 1991).

\bibitem{Andrieux2008b} Andrieux, D. \& Gaspard, P. Nonequilibrium generation of information in copolymerization processes. \textit{Proc. Natl. Acad. Sci. USA}, \textbf{105}, 9516-9521 (2008).

\bibitem{Pathria2011} Pathria, R.~K. \& Beale, P.~D. {\it Statistical Mechanics (Third Edition)} (Academic Press, Boston 2011).

\bibitem{SantaLucia2004} SantaLucia, J., Jr. \& Hicks, D. The thermodynamics of {DNA} structural motifs. \textit{Annu. Rev. Biophys. Biomol. Struct.} \textbf{33}, 415-440 (2004).

\bibitem{Breuer2012} Breuer, H.-P. Foundations and measures of quantum non-Markovianity. \textit{J. Phys. B: At. Mol. Opt. Phys.} \textbf{45} 154001 (2012). 

\bibitem{Rivas2014} Rivas, A., Huelga, S. F. \& Plenio, M. B. Quantum non-Markovianity: characterization, quantification and detection. \textit{Rep. Prog. Phys.}, \textbf{77}, 094001 (2014).\\ [+1mm]
\end{thebibliography}
\end{document}